\documentclass[useAMS]{gGAF2e}
\usepackage{graphicx}

\renewcommand*{\Psi}{\varPsi}
\renewcommand*{\Omega}{\varOmega}
\newcommand{\ephi} {{\bm {\hat e}}_\phi} 
\newcommand{\eal}{{\itshape{et al.}}}
\newcommand{\dd}{\mathrm d}

\def\rot{\nabla\times }
\def\p{\partial}
 
\def\eT{\eta_{t}}

\def\Om{{\it \Omega}}

\def\be{\begin{equation}}
\def\ee{\end{equation}}    
\def\ba{\begin{eqnarray}}
\def\ea{\end{eqnarray}}    

\begin{document}
\doi{}
\issn{} \issnp{} \jvol{00} \jnum{00} \jyear{2011} 
\markboth{An overshoot dynamo with a strong return flow}{}
\title{An overshoot solar dynamo with a strong return meridional flow}
\author{Alfio Bonanno\\
\vspace{6pt}
INAF, Osservatorio Astrofisico di Catania, Via S.Sofia 78, 95123 Catania, Italy}

\maketitle

\begin{abstract}
The meridional circulation plays an essential role in determining the basic mechanism of the dynamo action 
in case the of a low eddy diffusivity. Flux-transport dynamos with strong return flow and a deep stagnation point
are discussed in the case of  a positive $\alpha$-effect located in the overshoot layer and a rotation law consistent
with helioseismology. By means of a linear dynamo model, 
it will be shown that the migration of the toroidal belts at lower latitudes and the 
periods of the activity cycles are consistent with the observations. Moreover, at variance with previous investigations, 
the typical critical dynamo numbers of dipolar solutions are significantly smaller that those of quadrupolar solutions even 
in the regime of strong flow.
\end{abstract}


\section{Introduction}
The flux-transport dynamo is a promising mechanism to explain  several properties of the solar activity cycle. 
Although some aspects still need to be clarified, the basic observation behind this process is that 
in the presence of a low eddy diffusivity $\eT$ 
the magnetic Reynolds number ${L\, U}/{\eT}$ becomes  very large and the dynamics of the mean-field flow $U$ is thus an essential 
ingredient of the dynamo process. 
In this regime
the advection produced by the meridional circulation dominates  
the diffusion of the magnetic field which is then ``transported" by the meridional circulation. Clearly the word ``transport" must be used with some 
care because the field is not completely ``frozen" into the plasma: rather, the propagation of the dynamo wave is significantly distorted so that  
the well-known Parker-Yoshimura law is not necessarily  satisfied in this situation. 

Local helioseismology and doppler speed measurements substantially agree on a detection of
an average surface flow of about $15\; {\rm m\,s}^{-1}$ around $30^\circ$, with a peak
of about  $20-25\; {\rm m\,s}^{-1}$.   The return flow, located near the base of the convection zone,  
is much more difficult to detect and only theoretical estimates are at our disposal.  
In fact previous models of  flux-transport dynamo have considered a return flow of about one
order of magnitude smaller than the surface flow \citep{bo02, dikpati99, pij04,Gue08} although 
in recent studies \citep{manf08,manf11} it has been argued 
that the strength of this flow is instead of the same order 
of the poleward surface flow.

The other important ingredients of the model are the $\alpha$-effect and the differential rotation. 
While the latter can be determined by helioseismology, for the former
it will  be assumed that the most suitable location for this effect   \citep{Par55,skr66}
is just beneath the convection zone \citep{Par93} where a strong radial shear is produced in the so called tachocline \citep{SZ92}. 
The source of the turbulent helicity producing the $\alpha$-effect can be attributed to various mechanisms. The most promising one is the 
tachocline instability proposed by \cite{dikpati01}, although in recent investigations another appealing possibility is provided by 
a current-helicity generated $\alpha$-effect  \citep{gellert} due to kink and quasi-interchange instabilities 
in stably stratified plasmas \citep{bu11}.
The aim of this work is to study models of flux-transport dynamo with a strong meridional flow, where the stagnation point is 
self-consistently determined from the balance of the angular momentum in a turbulent plasma, following the self-consistent 
approach of \cite{durney00}. It will be shown that dipolar solutions are strongly favored and that, in the advection dominated regime,  the period is 
essentially determined by the strength of the return flow, 
at variance with the conclusion obtained by a recent investigation \citep{pipin11}.

\section{Basic Equations}
As is well known, the magnetic induction equation reads
\be\label{induction}
{\partial {\bm B} \over \partial t} = {\bm \rot} ({\bm U}
\times {\bm B} + \alpha {\bm B}) - {\bm \nabla} \times \left( \eT {\bm \nabla} \times {\bm B}\right),  
\ee 
where $\eT$ is the turbulent diffusivity.  Axisymmetry  implies that relative to spherical coordinates the magnetic field ${\bm B}$ 
and the mean flow field ${\bm V}$ are given by 
\be
{\bm B} = B_\phi(r,\theta,t)\ephi
+ {\bm \nabla}\times [ A(r,\theta,t) \ephi], \;\;\;\;\;\;\;\;
{\bm V} = {\bm u}(r,\theta) + r\sin\theta \Omega(r,\theta)\ephi,
\ee
where $B_\phi(r,\theta,\phi)\ephi$ and ${\bm \nabla} \times [A(r,\theta,t)\ephi]$ 
are the toroidal and poloidal components of the magnetic field respectively.
Moreover, the meridional circulation ${\bm u}(r,\theta)$
and differential rotation $\Omega(r,\theta)$ are the poloidal and toroidal components
of the global velocity flow field ${\bm V}$.
In particular the poloidal and toroidal components of (\ref{induction}) respectively determine 
\begin{subequations}
\label{ddd-coupled}
\begin{eqnarray}
{\partial A \over \partial t}
+\frac{1}{s}({\bm u}   {\bm \cdot} {\bm \nabla}) (sA)  &=& \alpha B +\frac{{\eT}}{r}\frac{\p^2 (rA)}{\p r^2}
+\frac{\eT}{r^2}\frac{\p }{\p \theta}\Bigl( \frac{1}{s}\frac{\p (sA)}{\p \theta}\Bigr), \\[2mm]
\frac{\p B}{\p t} + s\rho({\bm u} {\bm \cdot}  {\bm \nabla}) \Bigl(\frac{B}{s\rho}\Bigr)&= &
\frac{\p \Om }{\p r}\frac{\p (A\sin\theta)}{\p \theta}
-\frac{1}{r}\frac{\p \Omega}{\p \theta}
\frac{\partial (sA)}{\partial r}+\frac{1}{r} \frac{\p}{\p r}\Big({\eT}\frac{\p (rB)}{\p r} \Big)\nonumber\\
&&+\frac{\eT}{r^2}\frac{\p }{\p \theta}\Bigl(\frac{1}{s}\frac{\p (sB)}{\p \theta}\Bigr)-
\frac{1}{r}\frac{\p}{\p r} \Big (\alpha \frac{\p (rA)}{\p r} \Big)
-\frac{\p}{\p\theta}\Bigl(\frac{\alpha}{\sin\theta}\frac{\p (A\sin\theta)}{\p\theta}\Bigr),
\end{eqnarray}
\end{subequations}
where $s=r\sin\theta$. The $\alpha$-effect is always antisymmetric with respect to the equator, so that we write
\be
\alpha = \frac{1}{4}\, \alpha_0 \cos\theta\Big [1+{\rm erf}\Bigl(\frac{x-a_{1}}{d}\Bigr)\Big ]\Big [1-{\rm erf} \Bigl(\frac{x-a_{2}}{d}\Bigr)\Big ],
\label{aa}
\ee
where $\alpha_0$ is the amplitude of the $\alpha$-effect, $x=r/R_\odot$ is the fractional radius, 
$a_1$, $a_2$ and $d$ define the location and the thickness of the 
turbulent layer. In our investigation, we use the  values $a_1=0.68$, $a_2=0.72$ and $d=0.025$. 
It is reasonable to imagine that below the tachocline the turbulent diffusivity decreases by a few orders of magnitude
from the turbulent value attained in the bulk of the convection zone. We can conveniently represent this transition with 
the following functional form 
\be\label{eta}
\eta = \eta_c+\frac{1}{2}(\eta_t-\eta_c)\Bigl[1+{\rm erf}\Bigl(\frac{r-r_\eta}{d_\eta}\Bigr)\Bigr],
\ee
where 
$\eta_t$ is the eddy diffusivity,  $\eta_c$ the magnetic diffusivity beneath the
convection zone and $d_\eta$ represents the width of this transition. In particular, we use the values
$\eta_t/\eta_c=10^2$, $d=0.02$ and $r_\eta=0.71$.

The components of the meridional circulation can be represented with the help of a stream function 
$\Psi(r,\theta)=-\sin^2\theta\cos\theta\,\psi(r)$ 
so that 
\be
u_r= \frac{1}{r^2\rho\sin\theta} \frac{\partial \Psi}{\partial\theta}=  \frac{1-3\cos^2\theta}{\rho r^2} \; \psi(r),\;\;\;\;\
u_\theta= -\frac{1}{r\rho\sin\theta}\frac{\partial \Psi}{\partial r}=
\frac{\cos\theta \sin\theta} {\rho r} \; {\dd\psi(r)\over \dd r}
\label{13}
\ee
with the consequence that the condition $ {\bm \nabla} {\bm \cdot} (\rho {\bm u}) = 0$ is automatically fulfilled. 
A positive $\psi$ describes a cell circulating clockwise in the northern hemisphere, {i.e.} the flow is 
polewards at the bottom of the convection zone and equatorwards at the surface.
For a negative $\psi$ the flow is, as is observed, polewards at the surface. In 
order to keep the flow inside the convection zone, the function $\psi$ must be zero at 
the surface and at the bottom of the convection zone.
The  helioseismic profile for the differential rotation is taken so that 
\begin{equation}
\Omega(r,\theta)=\Omega_c+\frac{1}{2}\Bigl[1+\mathrm{erf}
  \Bigl(\frac{r-r_c}{d_c}\Bigr)\Bigr]
\bigl(\Omega_s(\theta)-\Omega_c\bigr), \quad 
\label{eq3}
\end{equation}
\noindent where $\Omega_c/2 \pi=432.8$ nHz is the uniform angular
velocity of the radiative core,
$\Omega_s(\theta)=\Omega_{eq}+a_2\cos^2  \theta+a_4\cos^4 \theta$
is the latitudinal differential rotation at
the surface. In particular $\Omega_{eq}/2\pi$$=$$460.7$ nHz is the angular
velocity at the equator,  $a_2/2\pi$$=$$-62.9$ nHz and
$a_4/2\pi$$=$$-67.13$ nHz, erf$(x)$ is the usual error function.
In this calculation the angular velocity is normalized in terms of equator differential 
rotation $\Omega_{eq}$, $r_c=0.71$ and $d_c=0.025$. 
As usual, the dynamo equations can be made dimensionless by introducing the dynamo numbers
\be
C_\Omega = \frac{R^2 \, \Omega_{\rm eq}}{\eT}, \;\;\;\;\; C_\alpha = \frac{R\,\alpha_0}{\eT}, \;\;\;\;\;
C_\omega = \frac{R \, \omega}{\eT}, \;\;\;\;\;\;\;\; C_u=\frac{R\, U}{\eT},
\ee
where $R$ is the stellar radius,  $\omega$ is the frequency of the dynamo wave and $U=u_\theta(r=R, \theta=45^\circ)$.
The meridional circulation is largely unknown, although helioseismology can provide an upper limit 
to the strength of the poleward flow as discussed in the introduction. 
A strategy to constraint several properties of the 
meridional circulation is to assume the differential rotation profile 
$\Omega(r,\theta)$ as a given ingredient, and deduce 
an approximation for the function $\psi$ 
from the angular momentum conservation along the azimuthal direction as shown by \cite{durney00}.
An approximate expression for $\psi$  is thus 
\be\label{dupsi}
\psi\approx \frac{5\rho r \tau }{2  } \int_0^\pi \langle u_r u_\theta \rangle {\mathrm d}\theta
\ee
in which $\tau$ the Coriolis number.
In particular, for the standard, isotropic mixing-length theory (\ref{dupsi}) becomes (see Durney 2000 for details)
\be\label{dupsi2}
\psi\approx - 5\rho r \tau \langle u_r^2\rangle\,.
\ee
In principle it would be possibile to explicitly compute $\psi$ and $\psi'$ using the relation 
(\ref{dupsi2}) knowing the convective velocities of the underlying stellar model.
\begin{figure}
\centering
\includegraphics[width=0.3\textwidth]{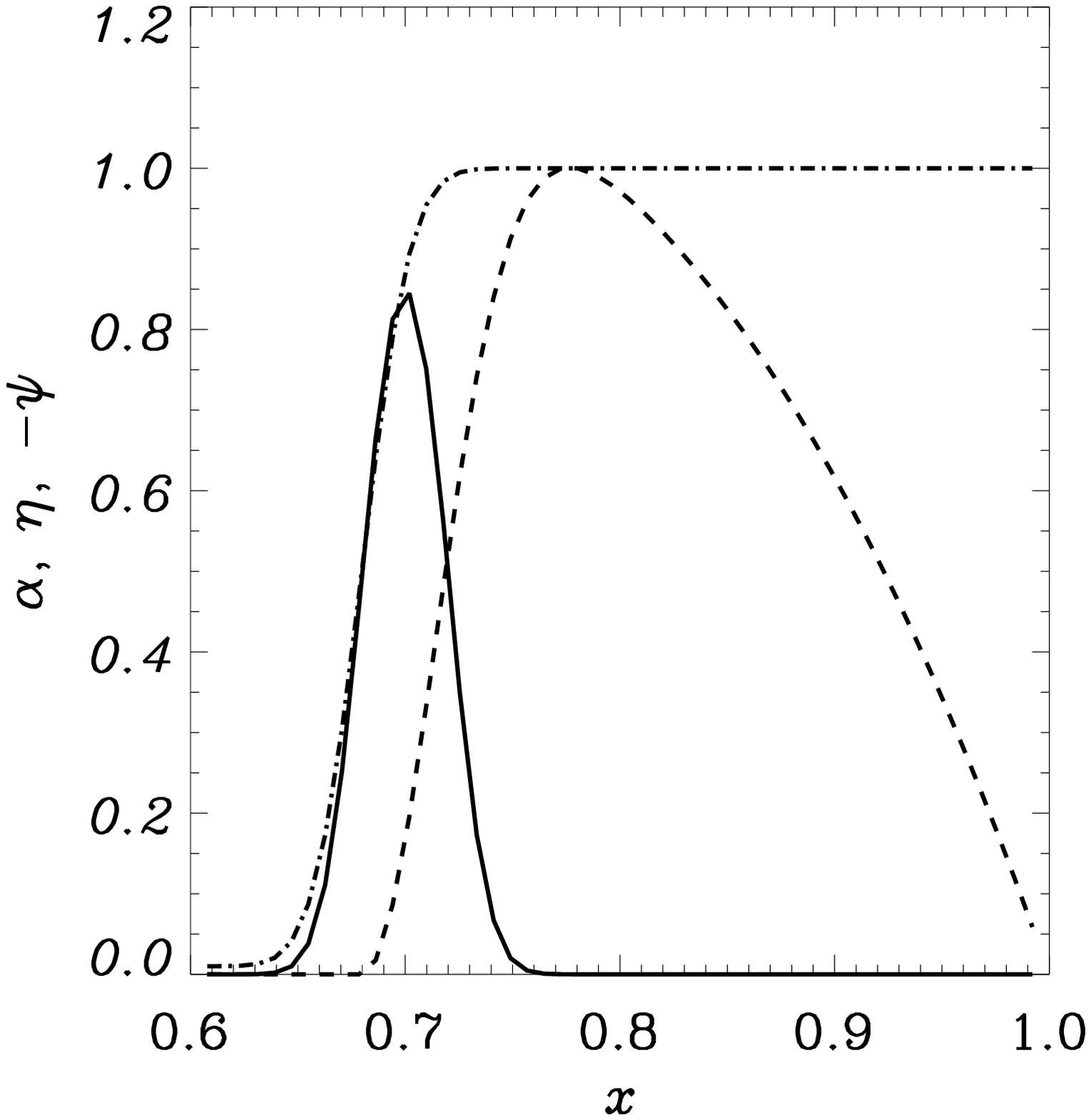}
\includegraphics[width=0.2\textwidth]{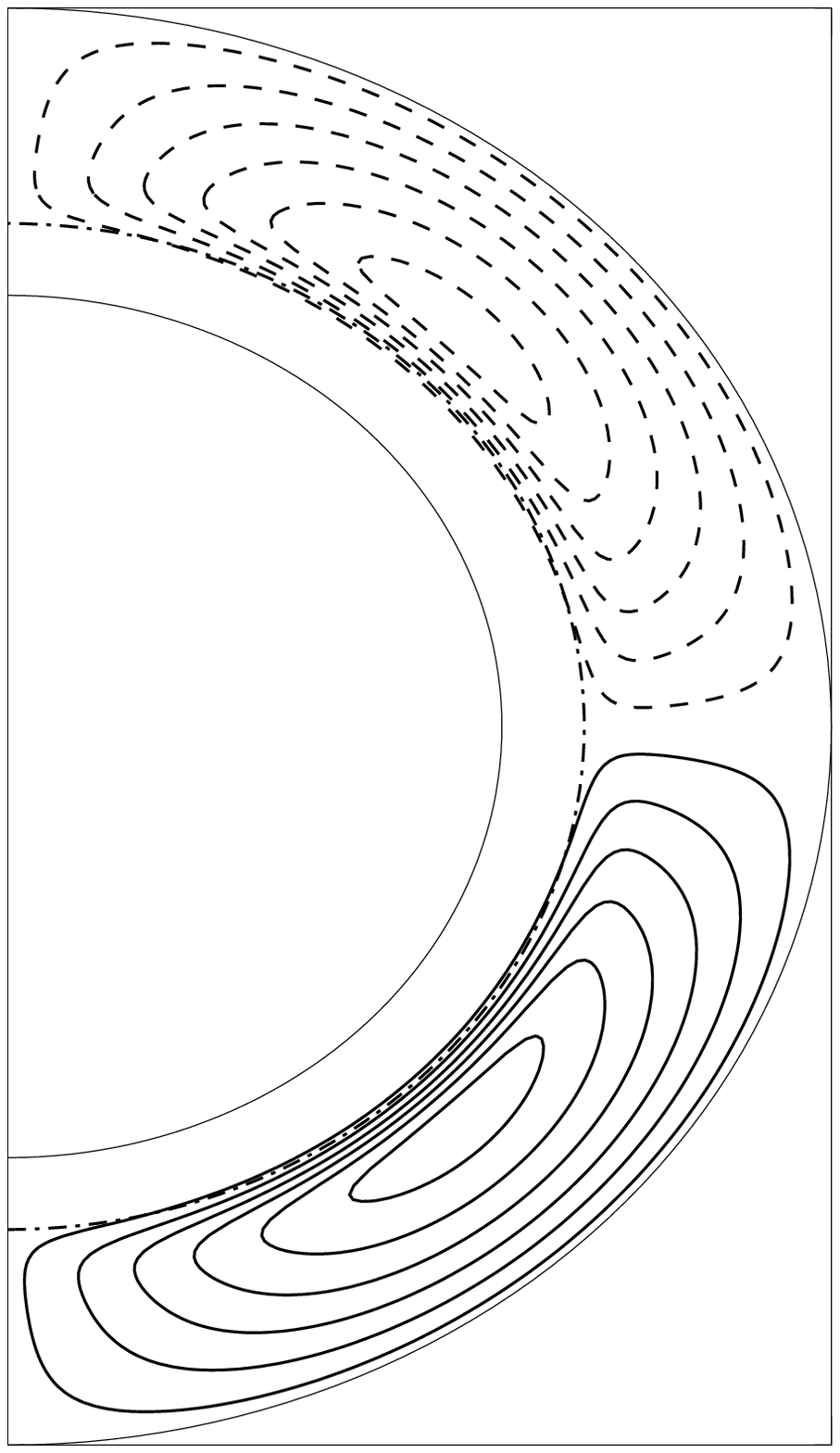}
\includegraphics[width=0.3\textwidth]{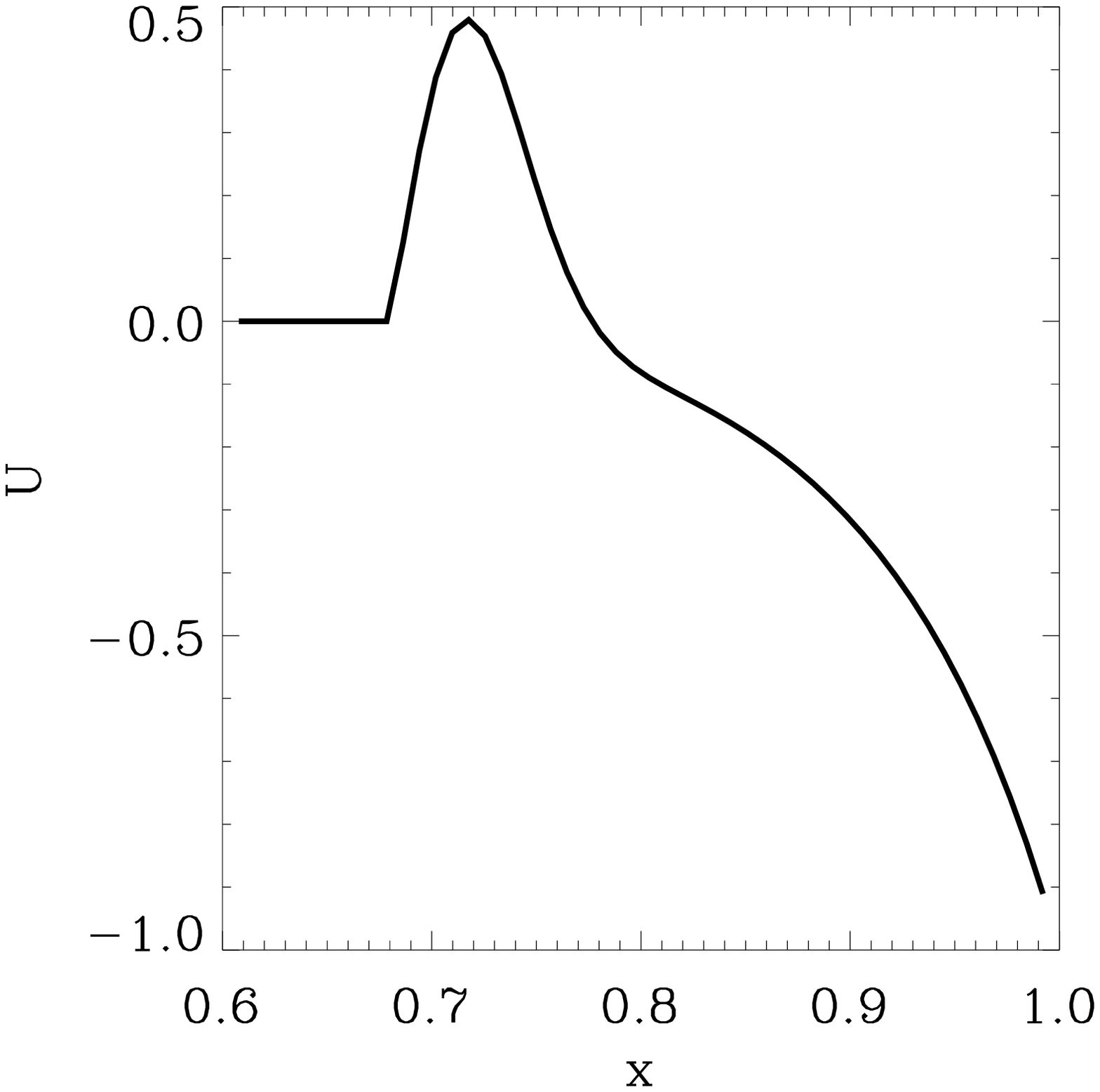}
\caption{The $\alpha$-effect (solid line), turbulent diffusivity (dot-dashed line) and (minus) the function $\psi(r)$ (dashed line), 
are depicted in the left panel. The streamlines of the flow are shown in the middle panel, dashed lines label  the anticlockwise flow.
The meridional circulation in units of the maximum surface value
at a latitude of $45^\circ$ is instead depicted in the right panel.}
\label{fig:stream}
\end{figure}
In practice this would be problematic, because the convective fluxes and their radial derivatives computed from standard MLT are 
discontinuous at the base of the convective zone. In a more realistic situation  the presence of an overshoot layer implies that
$\langle u_r^2\rangle \rightarrow 0$ smoothly
so that $u_\theta $ is continuous at the inner boundary. 
Nevertheless one can use the representation (\ref{dupsi2}) to determine the stagnation point where $\psi'=0$, which turns out to be
around $r=0.8$ solar radii in a standard solar model. 

This value is only slightly greater than the value obtained by \cite{manf11}, where the stagnation point is at $r=0.77$ (private communication). 
On the the other hand
those authors uses a stress-free boundary condition for the meridional circulation that implies a non-zero value of the flow at the inner boundary.
The problem is that in the overshoot region the eddy diffusivity drops of several orders of magnitude and the Reynolds number becomes very large
at the inner bottom,  most probably producing a boundary layer instability. I argue that stress free boundary conditions for the meridional circulation 
in the context of flux-transport dynamo are probably unphysical and one should try to consider a situation where both $u_r$ and $u_\theta$ vanishes
at the inner boundary.

An explicit form of the  function $\psi$ which incorporates the following features reads
\be
\psi=C \; \left[1-\exp{\left(-\frac{(x-x_b)^2}{\sigma^2}\right)} \right] (x-1) \; x^2 \,,
\ee
where $C$ is a normalization factor, $x_b=0.67$ defines the  penetration of the flow, $\sigma=0.025$ measures how fast 
$\langle u_r^2\rangle$ decays to zero in the overshoot layer and the location of the stagnation point.
The density profile is taken to be
\be
\rho=\rho_0 \Bigl(\frac{1}{x}-x_0\Bigr)^m
\ee
in which $m$ is an index representing the  the stratification of the underlying solar model, its value in the region of interest is approximately $2$, 
and $x_0=0.9$. The radial profile of the $\alpha$-effect, turbulent diffusivity, stream function and meridional circulation used in the 
calculation is depicted in figure (\ref{fig:stream}).

This linear dynamo problem  is solved with a finite-difference
scheme for the radial dependence and a polynomial expansion for the 
angular dependence.
In particular, the following expansions for the field are used:
\begin{subequations}
\label{4.1-4.2}
\begin{eqnarray}
 A(r,\theta)&= e^{\lambda t}& \sum\limits_{n} a_n(r)\ {\mathrm P}_n^{1}(\cos\theta), \\
 B(r,\theta)&=e^{\lambda t}& \sum\limits_{m} b_m(r)\ {\mathrm P}_m^{1}(\cos\theta),
\end{eqnarray}
\end{subequations}
where $\lambda$ is the (complex) eigenvalue so that ${\mathrm {Im}}\{\lambda\} = \omega$, the frequency of the dynamo wave, 
$n=1,3,5,\dots$ and $m=2,4,6,\dots$ for antisymmetric modes,
and $n\leftrightarrow m$ for symmetric modes.
Vacuum boundary conditions at the surface are then translated into 
\be 
{\dd a_n\over \dd x} + (n +1)\;a_n = b_m = 0.
\label{4.3}
\ee
In the interior at $x=x_{\rm i}=0.6$ we have instead the set 
\be
x{\dd b_m \over \dd x} +b_m = a_n = 0,
\label{4.4}
\ee
which imply perfect conductor boundary conditions.  

On substituting (\ref{4.1-4.2}a,b)
into (\ref{ddd-coupled}a,b)
 one obtains an infinite set of ordinary differential equations
that can be conveniently truncated in $n$ when the desired accuracy is 
achieved. The system is in fact solved by means of a second order accuracy 
finite difference scheme and the basic computational task is thus
to numerically compute eigenvalues and eigenvectors of a 
block-band diagonal real matrix of dimension
$M\times n$, $M$ being the number of mesh points and $n$ the number of harmonics,
$M(\alpha) v = \lambda v$
and $v$ is in general a complex eigenvector. This  algorithm is embedded in a bisection procedure in order to 
determine the critical  $C_\alpha$-value needed to find a purely oscillatory solution, for which 
${\mathrm {Re}}\{ \lambda\}=0$. For actual calculation the resolution of $60\times 60$ has been used because it was checked that further terms in the 
polynomial expansion and in the mesh points did not lead to any significant change in the solution.  
The code has  been extensively tested in \cite{jouve08}.

\section{Results}
\begin{figure}
\centering
\includegraphics[width=0.4\textwidth]{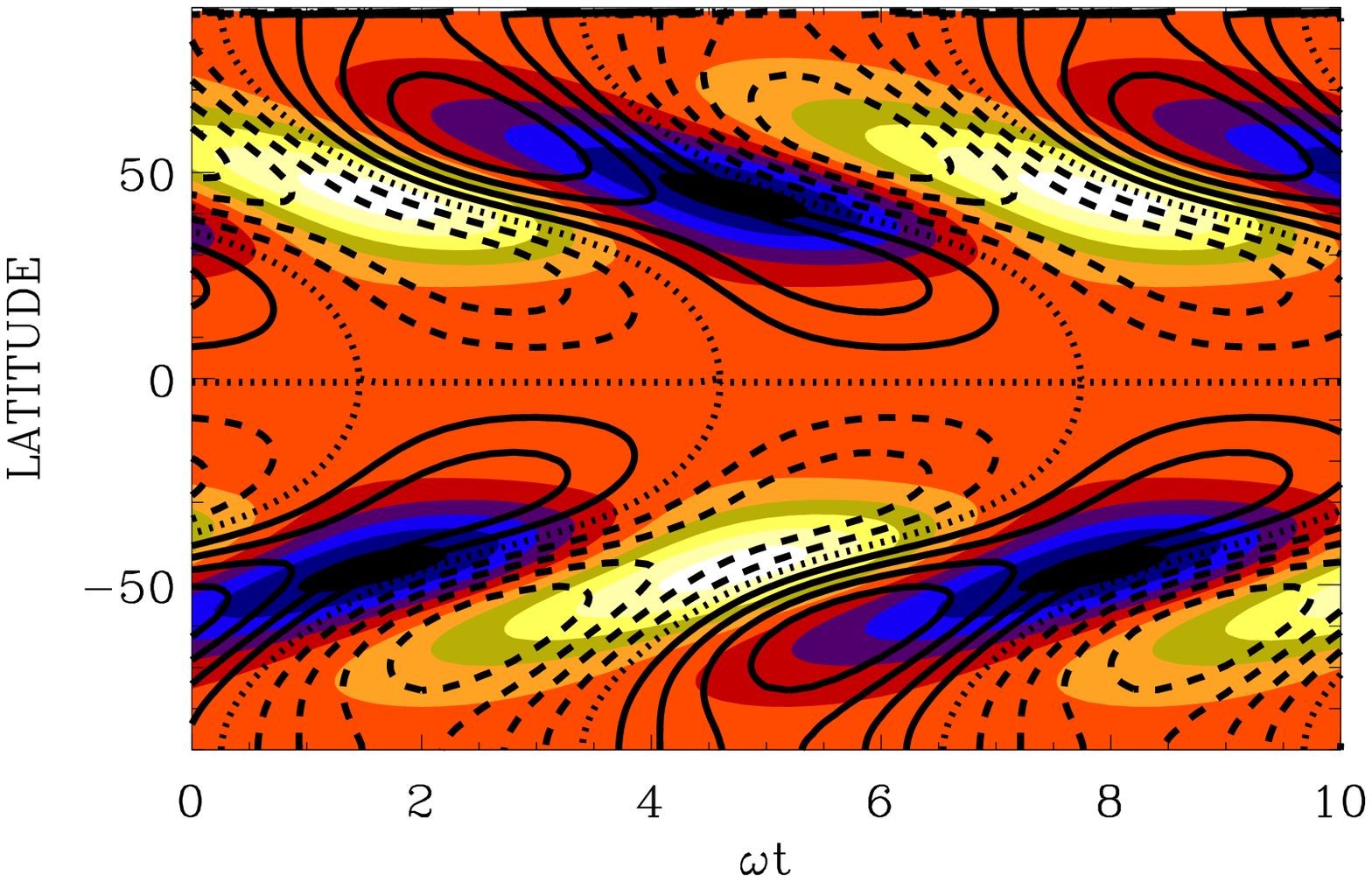} 
\includegraphics[width=0.4\textwidth]{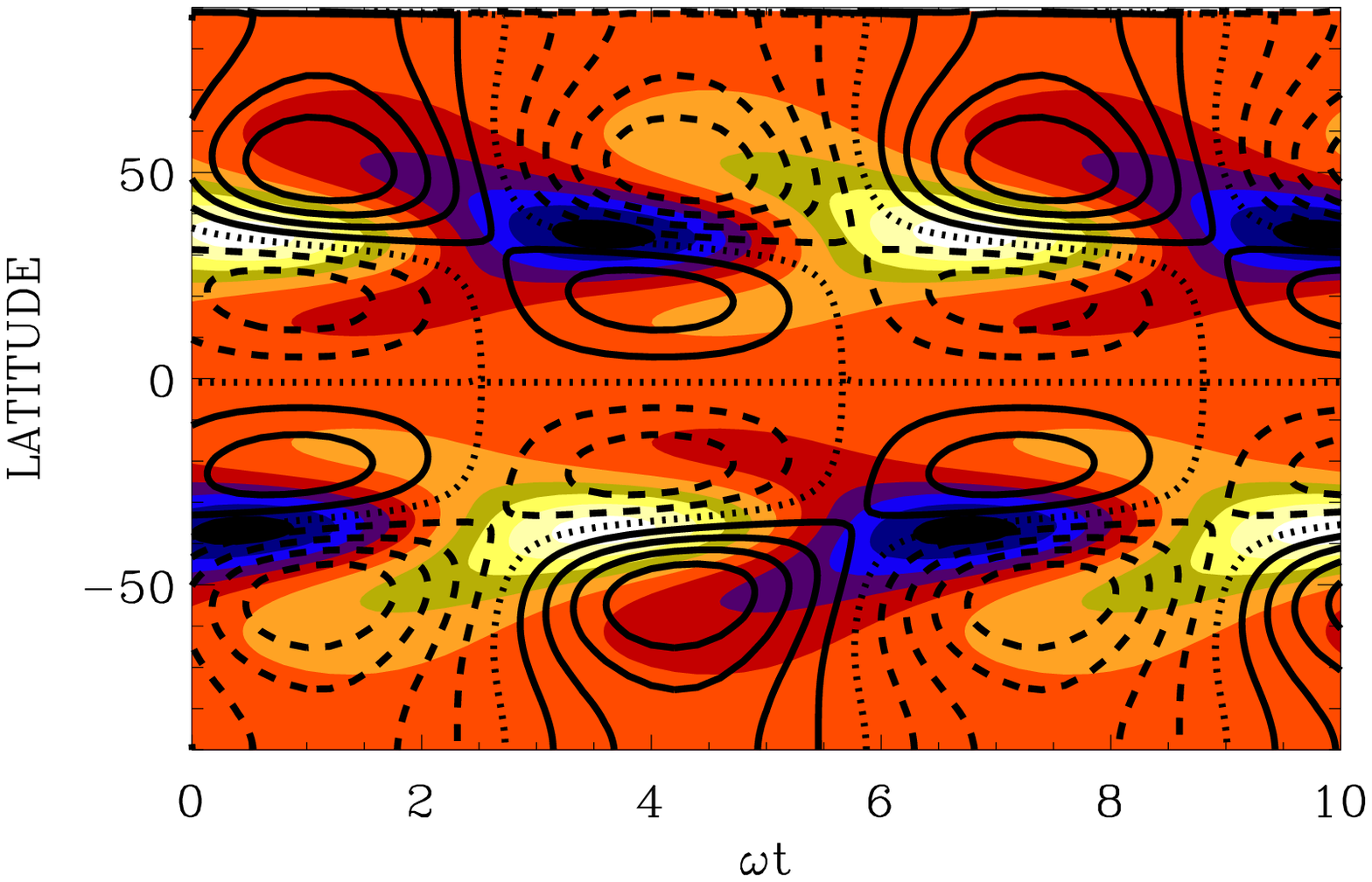}
\caption{Butterfly diagram for a overshoot dynamo solution with no meridional circulation and two different angular dependence
of the $\alpha$-effect. In the left panel the $\alpha$-effect is proportional to $\cos \theta$ and the solution has
$C_\alpha = 19.34$, $C_\Omega=3000$, and $C_\omega=33.22$. 
The solid and dashed lines represent the radial field at the surface  (solid for negative $B_r$), and blue is for negative toroidal field. 
The right panel represents instead a model with an $\alpha$-effect proportional to $\cos\theta \,\sin^2\theta$ with  $C_\alpha = 15.34$ and $C_\omega=23.02$. (colour online)}
\label{btflyov0}
\end{figure}

In this section the effect of a strong return flow with a deep stagnation point will be discussed. 
It is instructive to 
first consider the case of vanishing meridional circulation, where the propagation of the dynamo wave is
determined by the Parker-Yoshimura law.
In particular if the $\alpha$-effect has the simple  $\cos\theta$ angular dependence 
the generation of the poloidal field from the toroidal 
is more pronounced at higher latitudes  and, as $\partial \Omega / \partial r < 0$ there, 
the dynamo wave propagates towards the
equator until it reaches the latitude where $\partial \Omega / \partial r \approx 0$ and the migration is halted.
The solution is characterized by the following dynamo numbers, 
$C_\alpha = 19.34$, $C_\Omega=3000$, and frequency 
$C_\omega=33.22$, which implies a period of about $6.2$ years for the Sun given a turbulent diffusivity of  
$\eT=4.66 \; 10^{12} {\rm cm^2 s^{-1}}$. Models with a negative $\alpha$-term at the bottom of the convection zone
produce only stationary solutions.  The butterfly diagram is depicted in the left panel of figure (\ref{btflyov0}),
where the radius is taken at maximum $B_\phi$.
It is interesting to see what happens for an angular dependence of the type $\cos \theta \sin^2\theta$ for the $\alpha$-term since 
the field regeneration due to the dynamo action occurs at lower latitude in this case. The solution for a positive $\alpha$-term
is depicted in the right panel of figure (\ref{btflyov0}). In this case the dynamo action occurs around latitudes at which
$\partial \Omega / \partial r \approx 0$ and 
two distinct branches, one equatorward caused by the $\partial \Omega / \partial r < 0$ at high latitudes 
and the other  poleward, at lower latitude, caused by $\partial \Omega / \partial r < 0$ are present.
The poleward migration of the butterfly diagram at low latitude and the phase relation are clearly not consistent with the observations.
In the case of a negative $\alpha$-term the only possible solution is a stationary one, as before.
The conclusion of the above calculations is that it is difficult to imagine that a simple overshoot 
dynamo can successfully reproduce the observed features of the solar cycle, difficulties  also noticed in 
earlier investigations \citep{bra95}.

\begin{figure}
\centering
\includegraphics[width=0.3\textwidth]{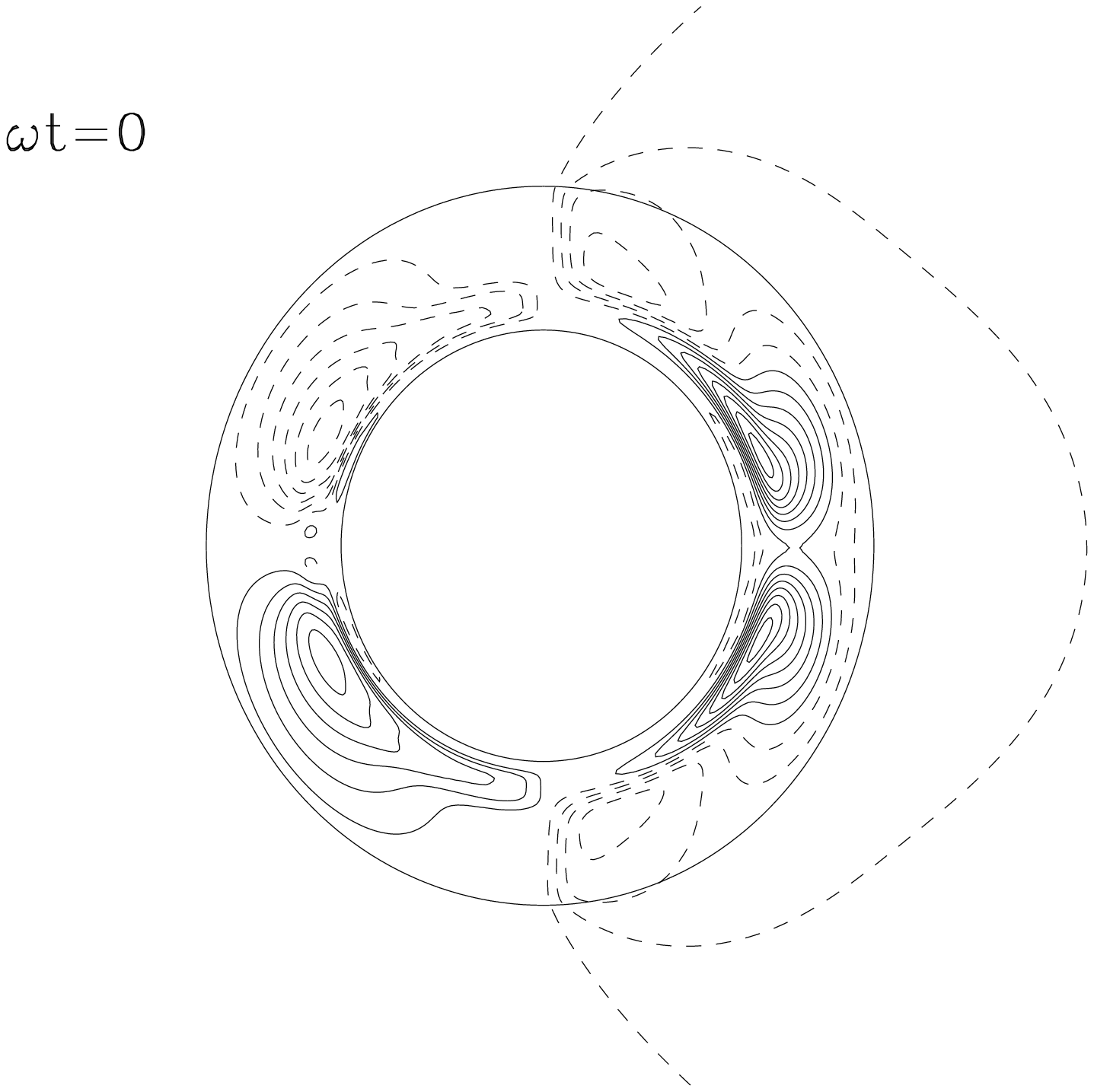}
\includegraphics[width=0.3\textwidth]{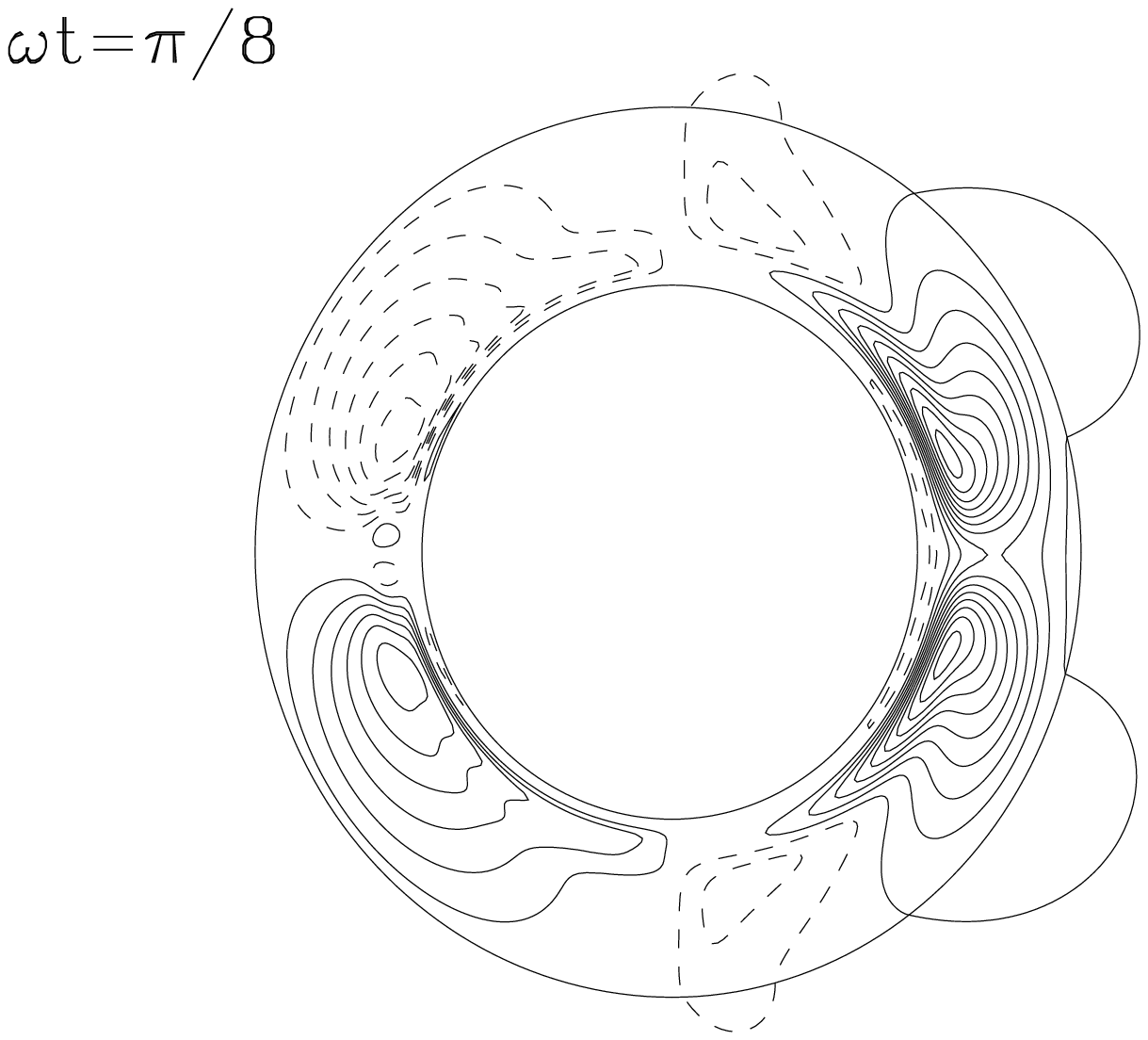}
\includegraphics[width=0.3\textwidth]{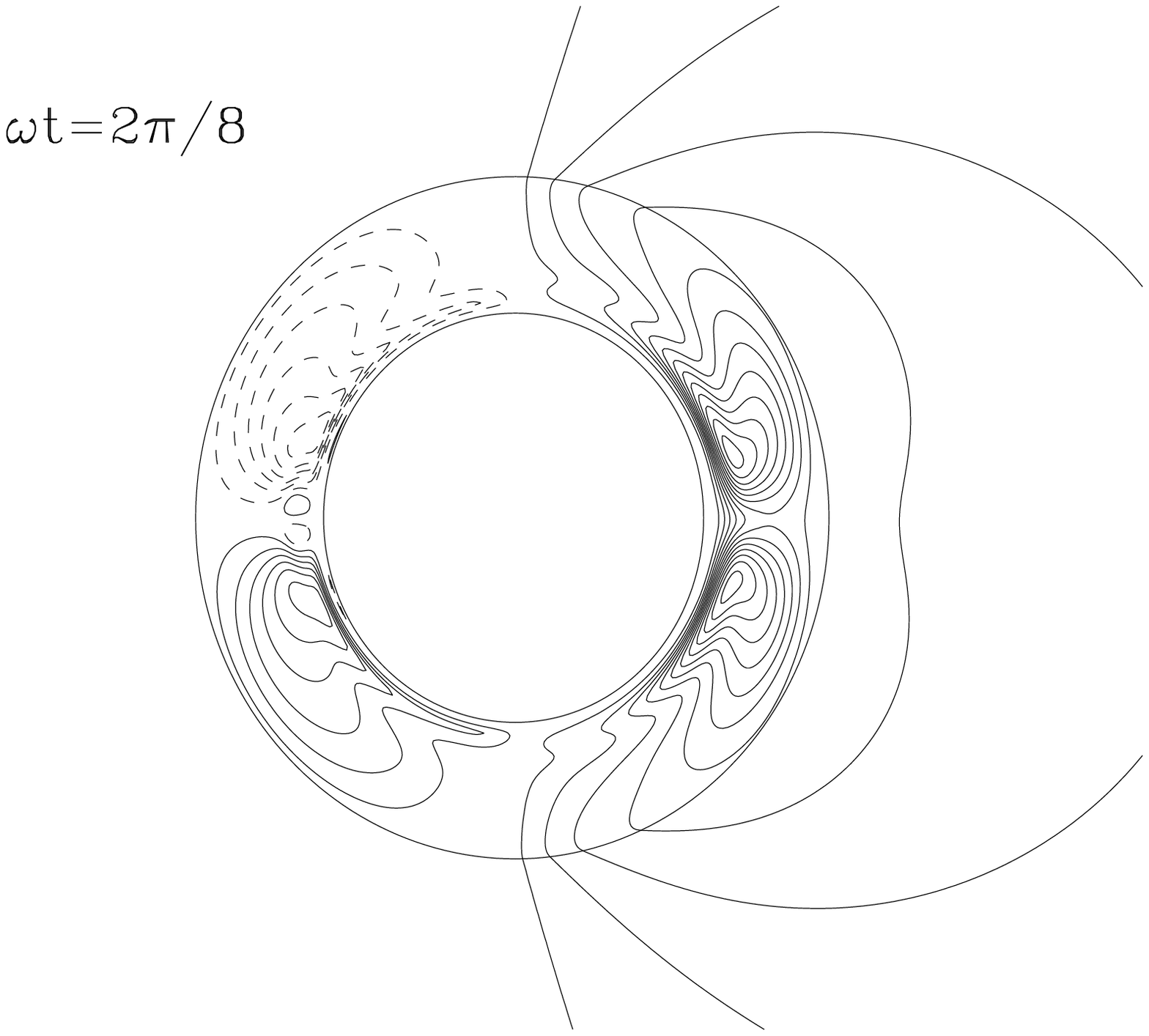}
\includegraphics[width=0.3\textwidth]{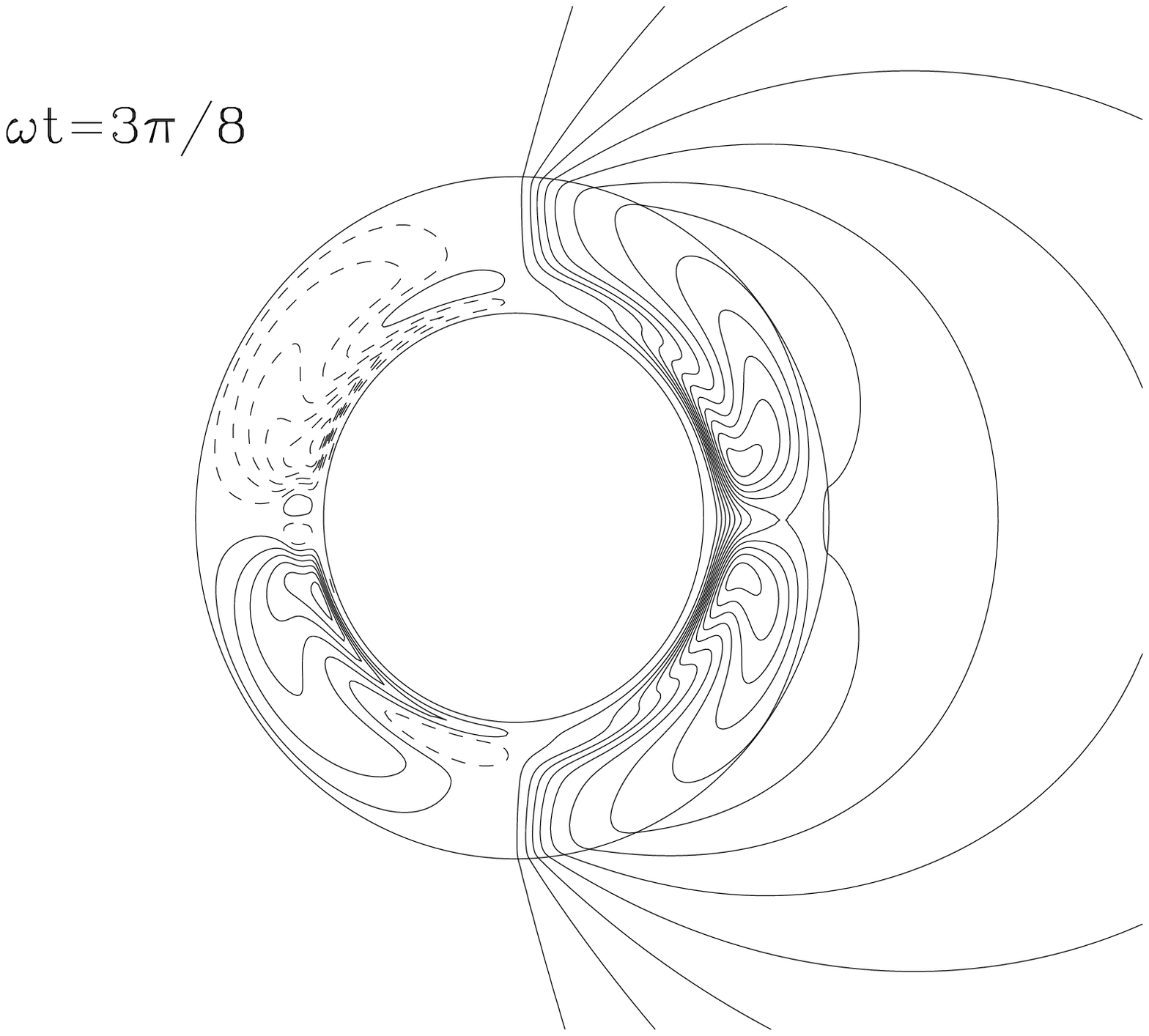}
\includegraphics[width=0.3\textwidth]{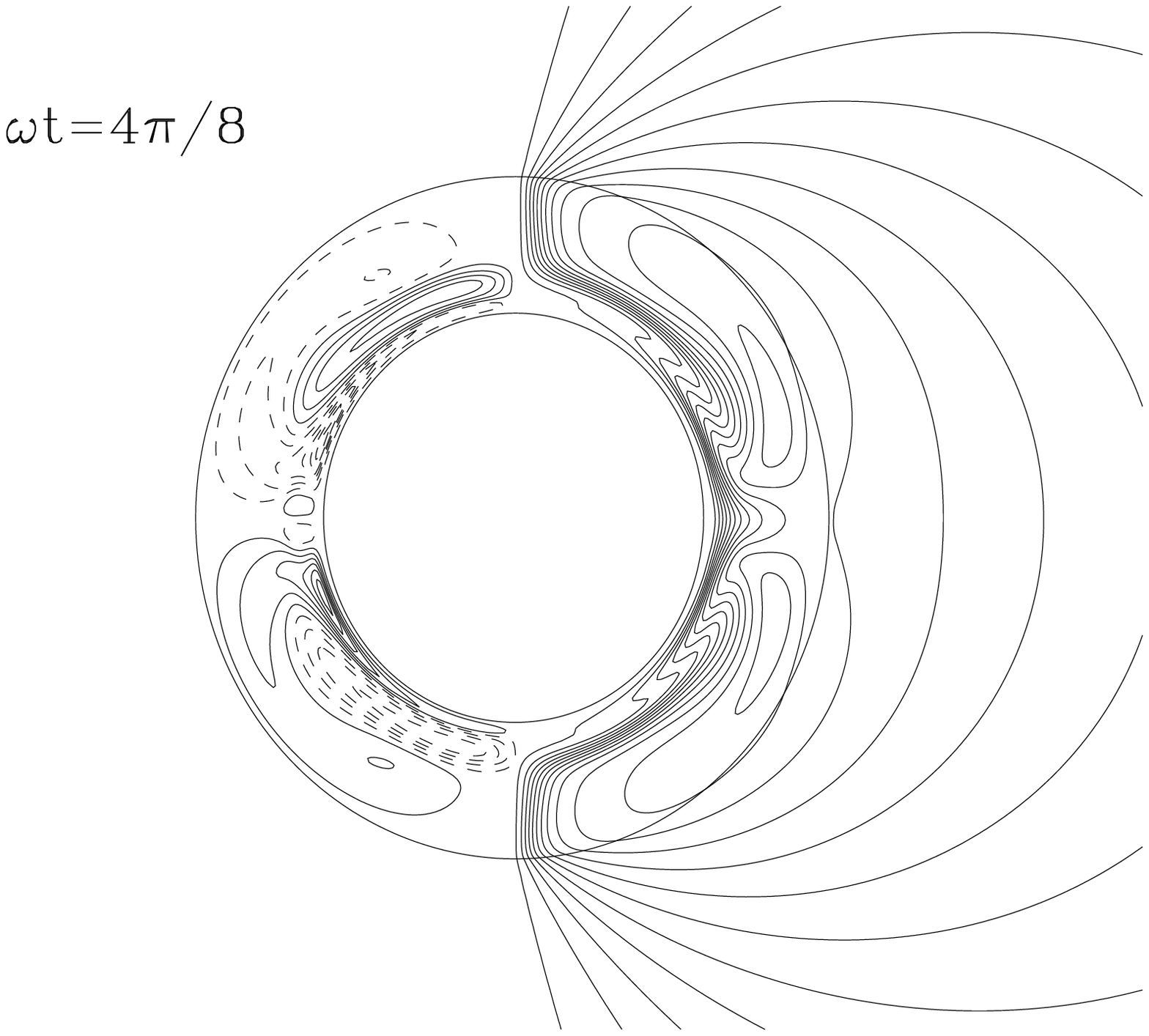}
\includegraphics[width=0.3\textwidth]{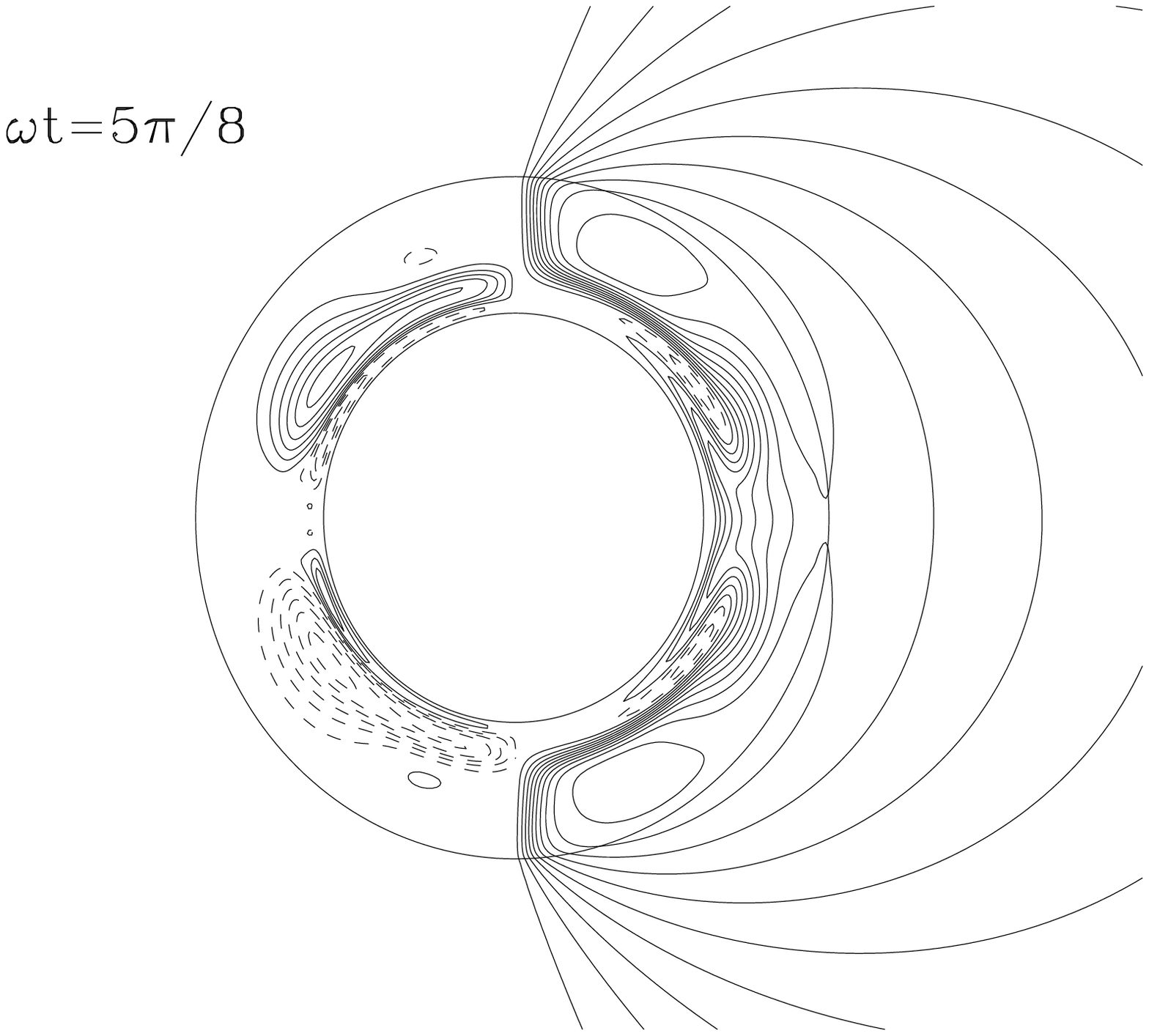}
\includegraphics[width=0.3\textwidth]{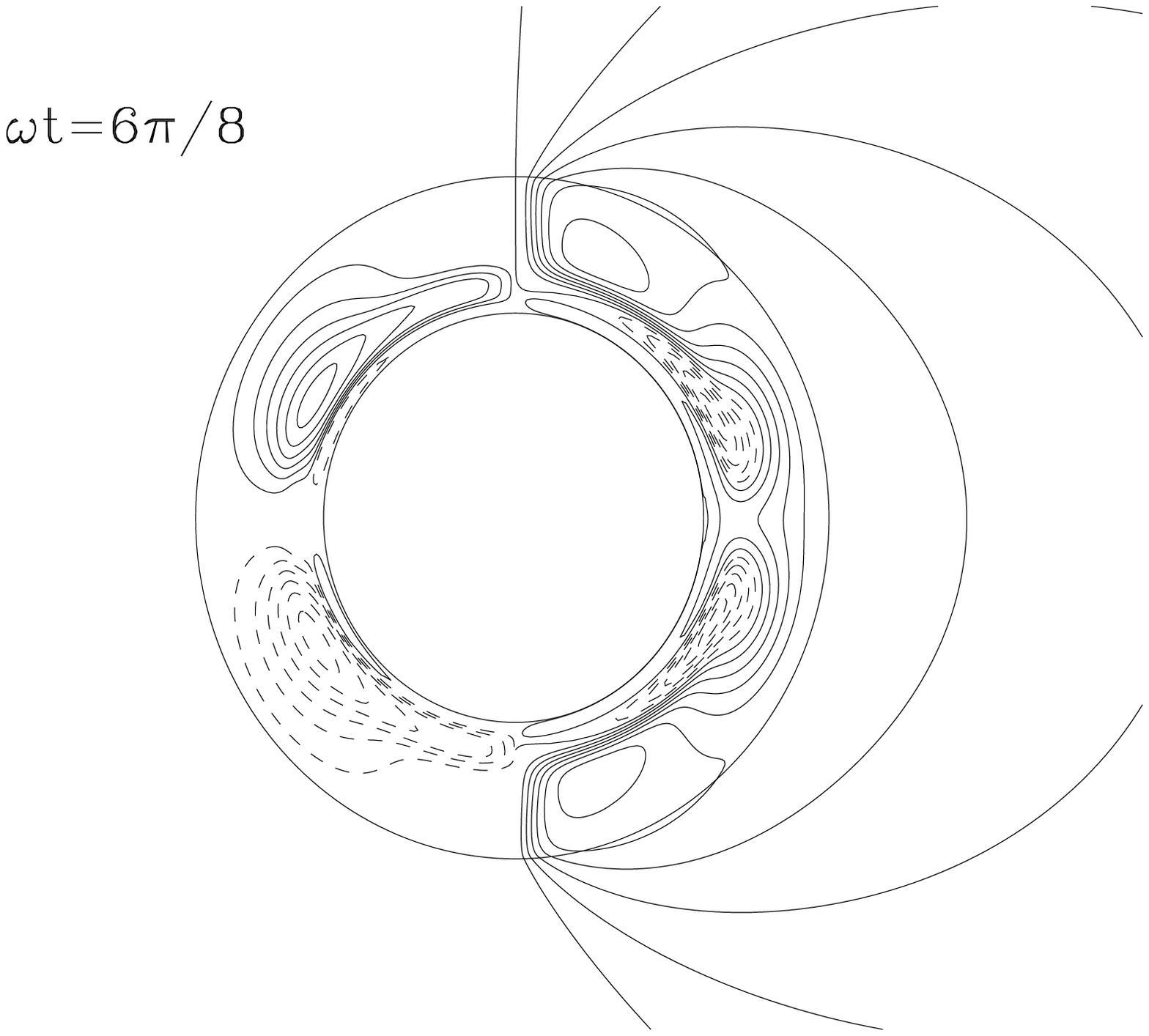}
\includegraphics[width=0.3\textwidth]{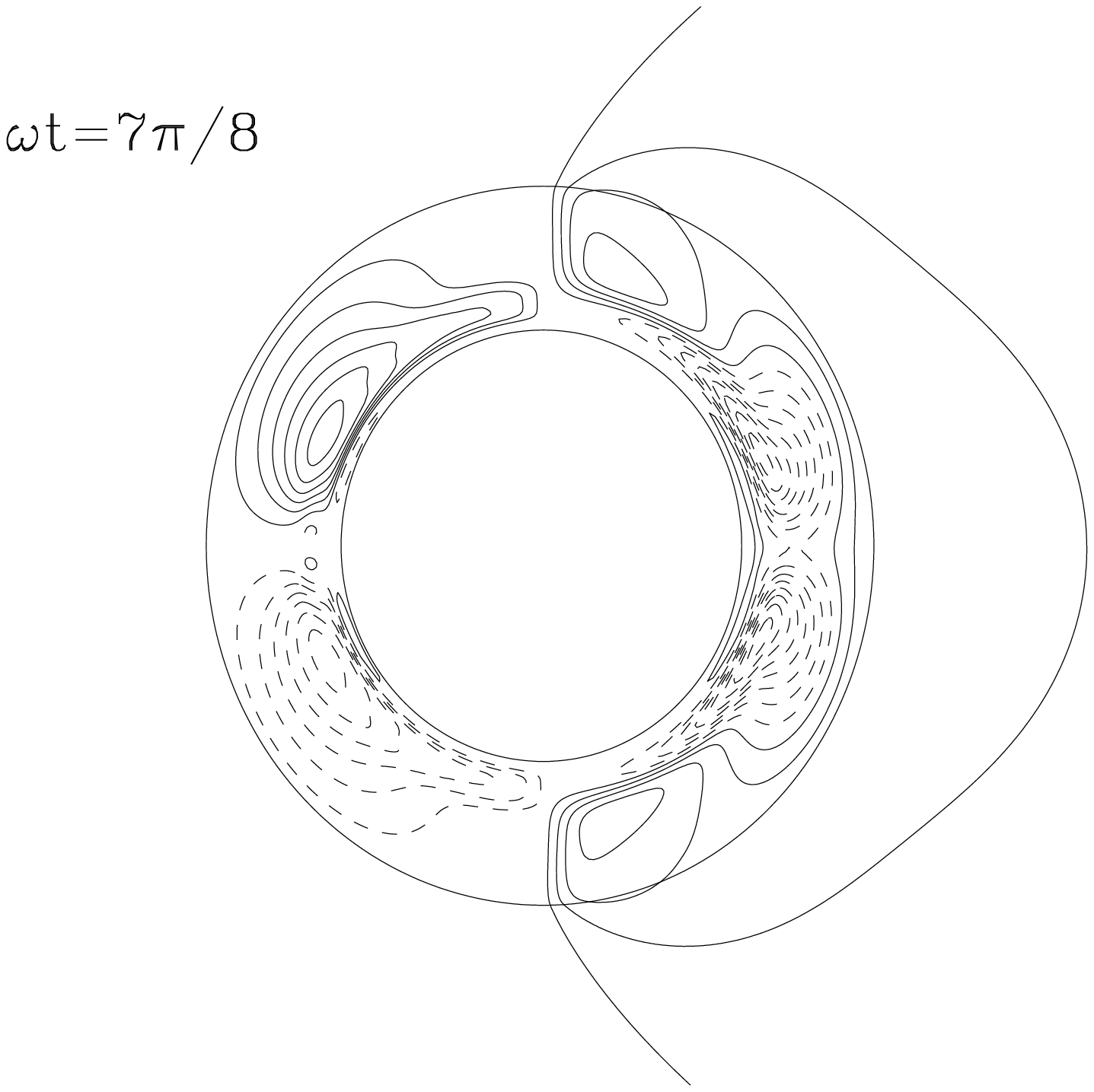}
\caption{The temporal evolution of the magnetic field for a solution with $C_u=400$,  
$C_\alpha = 19.34$, $C_\Omega=30000$, and frequency $C_\omega=102$ which imply a period of 
about $20$ yrs and a flow of $27 \, {\rm m\, s^{-1}}$. 
The left part are the isocontours line of the toroidal field, 
with solid line for negative $B_\phi$ and dashed line for positive value of the field. The right part represents of the streamlines of the poloidal field
given by contours of $A r \sin \theta$. Solid line are for negative values of $A$.}
\label{ov0m}
\end{figure}

The situation radically changes in the flux-dominated regime, $C_u\gg 1$ as one can see in figure (\ref{ov0m}) 
as in this case the meridional circulation transports the flux at lower latitudes. This solution is characterized by 
a flow $U=27 \, {\rm m\, s^{-1}}$ with $\eT=4.66 \; 10^{11} {\rm cm^2 s^{-1}}$ so that $C_u=400$.
\begin{figure}
\centering
\includegraphics[width=0.6\textwidth]{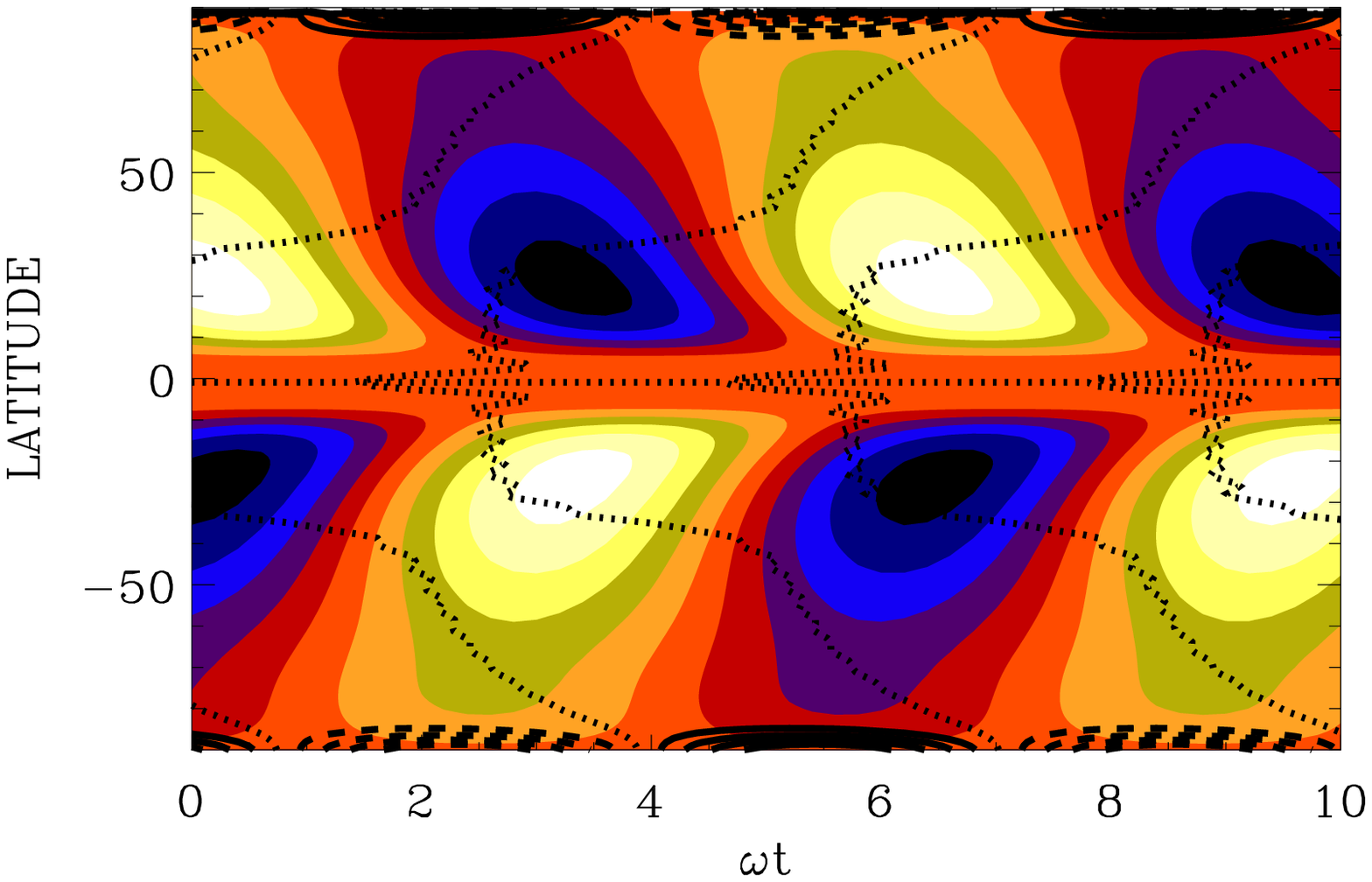} 
\label{btflyov0m}
\caption{Butterfly diagram and phase relation for the solution presented in figure (\ref{ov0m}). (colour online)}
\end{figure}
The relevant question is to discuss the parity of the solutions as a function of the strength of the meridional circulation since
a strong equatorward flow at the bottom of the convection zone can easily favor a quadrupolar parity. In fact in this case 
the field is  by definition non-zero at the equator and the presence of the meridional circulation can help the occurrence
of solutions with this symmetry. Usually only a small window in the parameter space is consistent with the observations \citep{bo06}.

\begin{figure}
\centering
\includegraphics[width=0.4\textwidth]{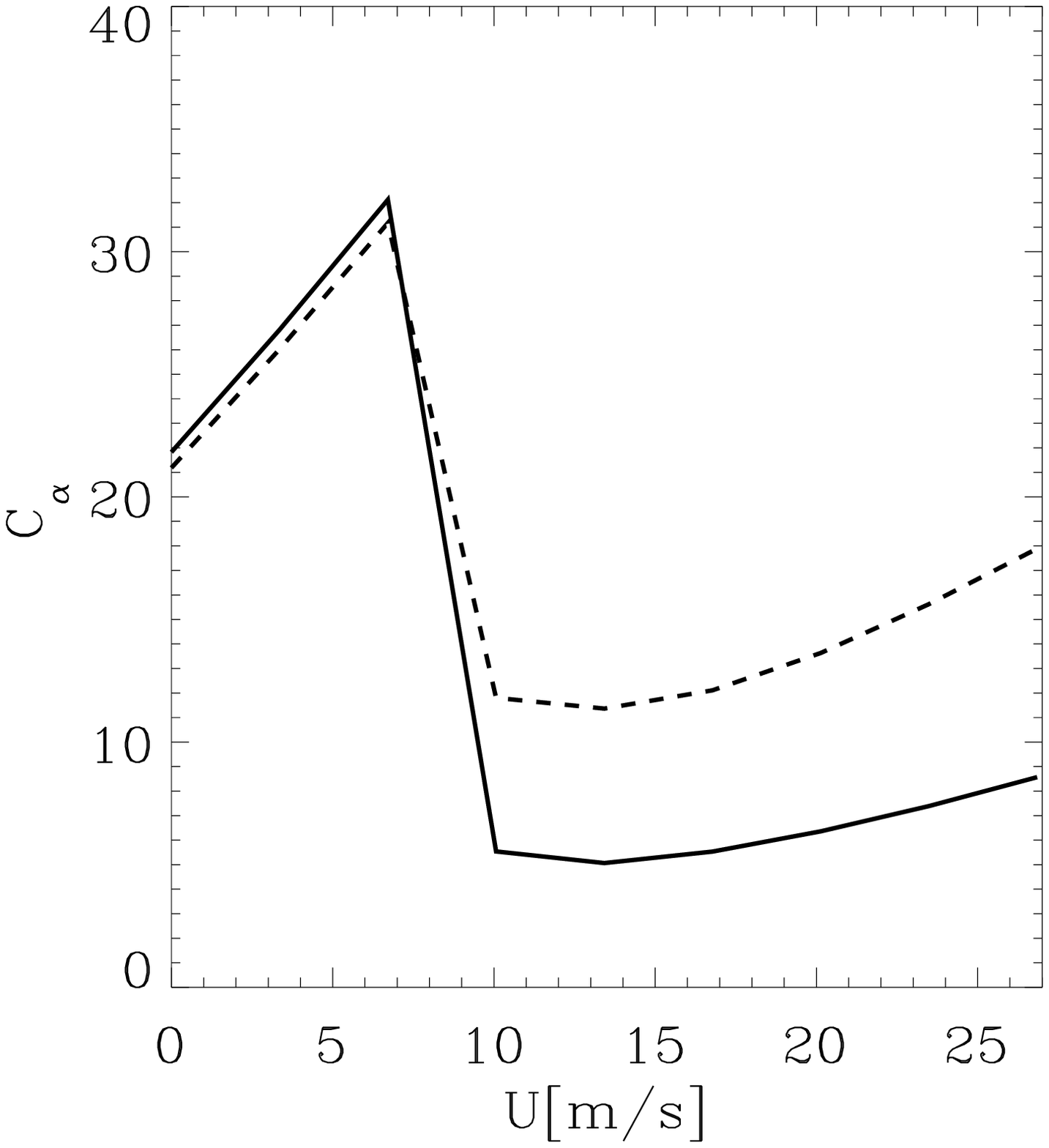}
\includegraphics[width=0.4\textwidth]{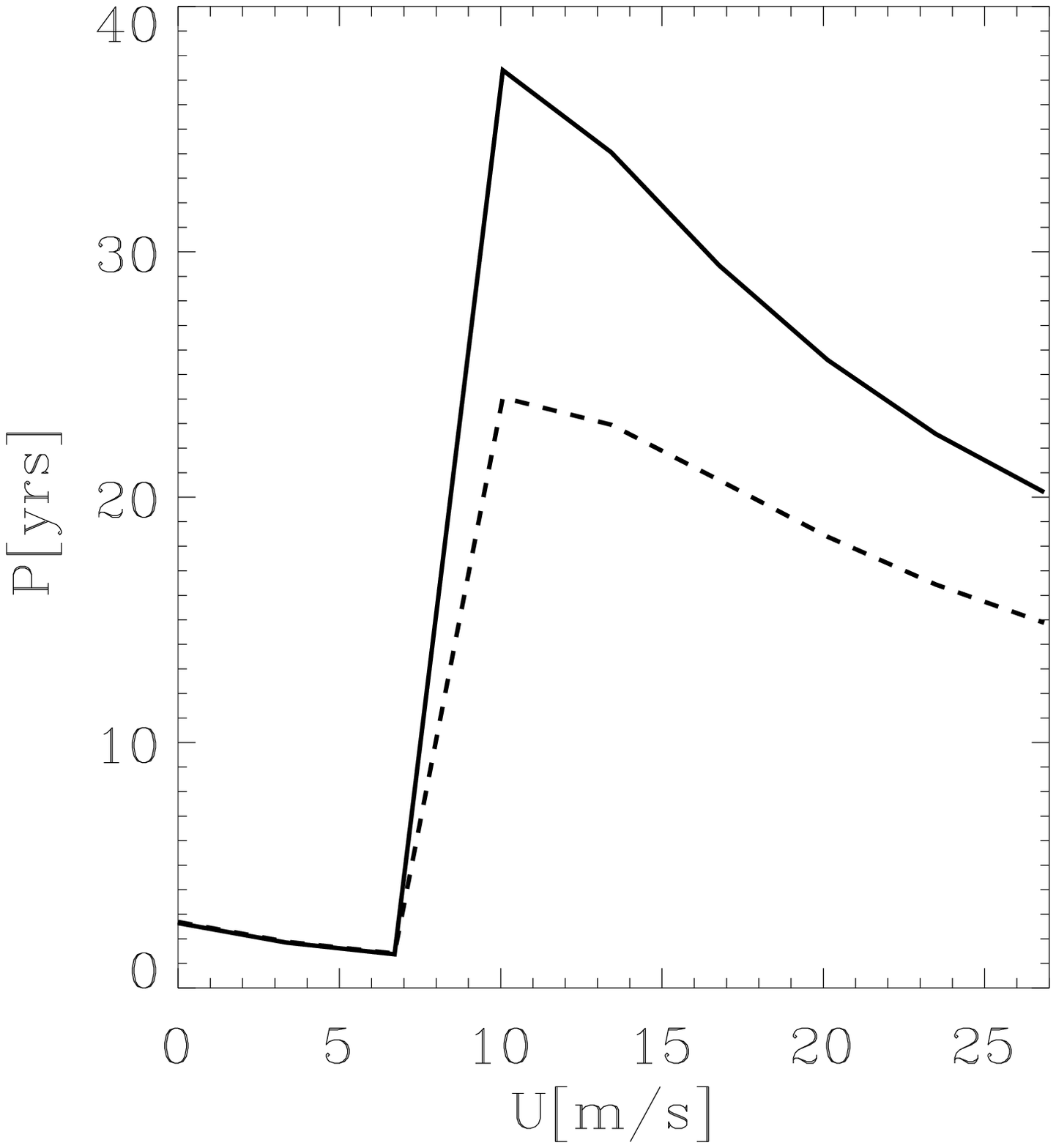}
\caption{The critical dynamo numbers (left panel) and the period (right panel)  of the solutions as a function of the meridional circulation
for dipolar solutions (solid line) and quadrupolar solution (dashed lines) for an eddy diffusivity of $\eT=4.66 \; 10^{11} {\rm cm^2 s^{-1}}$}
\label{parity}
\end{figure}

However, as it is shown in figure (\ref{parity}) the critical $C_\alpha$ value is significantly smaller for dipolar solutions (solid line)
than for quadrupolar solutions (dashed line), at variance with previous investigations where
the difference in the critical dynamo numbers between dipolar and quadrupolar solution was not found to be very significative \citep{bo02}.
It is important to stress that the evolution of the critical dynamo number $C_\alpha$ and the period {\it is not} a monotonic function
of the meridional circulation, in general. As it is clear from figure (\ref{parity}) there are essentially two different scaling regimes 
for weak and for strong flow. In the first case the meridional circulation does not significantly affect the behavior of the dynamo wave, 
while at large value of the flow, the period decreases as the flow increases.
The two scaling regimes are separated by a region of stationary solution around $7 \, {\rm m s^{-1}}$ for an eddy diffusivity of 
$\eT=4.66 \; 10^{11} {\rm cm^2 s^{-1}}$.
This result is in agreement with the findings
of previous studies \citep{bo02, bo06} and it is not changed by the presence of a strong flow with a deep stagnation point.
I conclude that the difference with the findings of \cite{pipin11} are mainly due to the inclusion (in that work) of the ${\bm \varOmega} \times {\bm J}$  contribution 
in the turbulent electromotive force (see their figure 5 where a saturation of the period seems to occur instead).

\section{Conclusions}
The results presented in this investigation suggest that a standard $\alpha^2\Omega$ mean field dynamo
with a deep meridional circulation and an $\alpha$-effect located
in the overshoot layer reproduces the migration of the toroidal belts at lower latitude in dynamo models with heliosesimically derived 
rotation law. At variance with previous studies (with the exception of the recent work by Pipin \& Kosovichev 2011), the return flow is of the same
order of the surface flow, and the solutions are clearly dominated by models with dipolar parity. 

In this investigation the supercritical dynamo action has not been considered. This point was investigated (for a different flow pattern)
in \cite{jouve08} where it was shown that values of $C_\alpha$ about ten times supercritical does not significantly 
change the nature of the solution, even in the flux-transport regime.  On the other hand, the difference between dipolar and quadrupolar
solutions in the flux-dominated regime are rather significant, and one could argue that at least for not too large supercritical dynamo numbers the 
conclusions of this investigations will still be valid.

The  idea of determining the stagnation
point of the flow by the maximum of the convective fluxes in the convection zone, as suggested by \cite{durney00}, leads to 
successful models of the 
meridional circulation in flux-dominated dynamo. It would be important to extend the findings of this work 
to the stellar case, at least for slowly rotating solar-like stars.
\newcommand{\yanaS}[5]{, ``#5'' {\em Astron.\ Astrophys.\ }{\bf #2}, #3-#4 (#1).}
\newcommand{\yana}[5]{, ``#5,'' {\em Astron.\ Astrophys.\ }{\bf #2}, #3-#4 (#1).}
\newcommand{\yanc}[5]{, ``#5'' {\em Astron.\ Nach.\ }{\bf #2}, #3-#4 (#1).}
\newcommand{\yapj}[5]{, ``#5,'' {\em Astrophys.\ J.\ }{\bf #2}, #3-#4 (#1).}
\newcommand{\yjfm}[5]{, ``#5,'' {\em J.\ Fluid Mech.\ }{\bf #2}, #3-#4 (#1).}
\newcommand{\ymn}[5]{, ``#5,'' {\em Monthly Notices Roy.\ Astron.\ Soc.\ }{\bf #2}, #3-#4 (#1).}
\newcommand{\ysph}[5]{, ``#5,'' {\em Solar Phys.\ }{\bf #2}, #3-#4 (#1).}
\newcommand{\ypre}[5]{, ``#5,'' {\em Phys.\ Rev.\ E }{\bf #2}, #3-#4 (#1).}
\newcommand{\yprlN}[5]{, ``#4,'' {\em Phys.\ Rev.\ Lett. }{\bf #2}, #3 (#1).}
\newcommand{\yjourN}[5]{, ``#5,'' {\em #2} {\bf #3}, #4 (#1).}
\newcommand{\adsp}[5]{, ``#5,'' {\em Advances in Space Research} {\bf #2}, #3-#4 (#1).}
\newcommand{\jpcs}[5]{, ``#5,'' {\em Journal of Physics: Conference Series} {\bf #2}, #3-#4 (#1).}
\newcommand{\yzna}[5]{, ``#5,'' {\em Zeitschrift f{\"{u}}r Naturforschung} {\bf #2}, #3-#4 (#1).}

\end{document}